\documentclass[prb,floatfix,preprint]{revtex4}
\usepackage[final]{graphicx}
\usepackage[usenames]{color}
\usepackage{afterpage}

\begin{document}

\title{Anisotropic Exchange Interaction between Non-magnetic Europium Cations in Eu$_2$O$_3$}
\author{  G. Concas,$^{1,*}$,  J. K. Dewhurst, $^2$ A. Sanna, $^2$ S. Sharma$^2$, and S. Massidda$^3$}
\affiliation{$^1$Dipartimento di Fisica, 
Universit\`a degli Studi di Cagliari, and INSTM, S.P. Monserrato-Sestu km 0.700, I-09042 Monserrato (CA), Italy}
\affiliation{$^2$Max-Planck-Institut f{\"u}r Mikrostrukturphysik, Weinberg 2, D-06120 Halle, Germany}
\affiliation{$^3$CNR-IOM SLACS, and Dipartimento di Fisica, 
Universit\`a degli Studi di Cagliari, S.P. Monserrato-Sestu km 0.700, I-09042 Monserrato (CA), Italy}
\email{giorgio.concas@dsf.unica.it}

\begin{abstract}

The electronic structure of the cubic and (high pressure)  hexagonal phases of
Eu$_{2}$O$_{3}$ have been investigated by mean of full potential linearized augmented plane wave calculations, 
within the LDA+$U$ method. 
For the cubic phase, the comparison between ferromagnetic and antiferromagnetic calculations shows that the exchange interaction is very weak and is therefore expected to have  a negligible effect on the magnetic susceptibility. This  is consistent with the experimental behavior of the susceptibility of solid solutions of Eu$_{2}$O$_{3}$ into A$_{2}$O$_{3}$ (A=Y, Lu, Sc).
The calculations  performed for the high pressure hexagonal phase, on the other hand, show that there is an antiferromagnetic exchange interaction between nearest neighbor Eu ions, which should have a sizeable effect on the susceptibility.  Our results allow us to discuss the existing theories.
\end{abstract}
\maketitle

\section{introduction}

Rare earth compounds present a large variety of interesting magnetic behaviors, due to
their partially filled $f-$shells. 
Trivalent europium compounds, in particular, offer an unique chance of observing 
exchange coupling between "non-magnetic" ions.
The ground state of Eu$^{3+}$ ($4f^{6}$) ions is $^{7}F_{0}$, with a
total angular moment $J=0$  resulting from both $L=S=3$ (in atomic units),
however,  spin-dependent exchange effects are present. 
Furthermore, Eu ions   show a substantial admixture of
higher energy $J=1$ states, which contributes significantly to their susceptibility ($\chi$).  

Eu$_2$O$_3$ is the prototypical compound of this family, its
magnetic susceptibility has been experimentally investigated 
in detail\cite{borovik56,antic97,concas08}. The first attempt to explain its magnetic behaviour 
 was made by Huang and Van Vleck\cite{huang69}, who showed that the susceptibility of Eu$_2$O$_3$ is larger than the corresponding susceptibility of the free Eu$^{3+}$ ion, because the energy levels of the excited $^7$F$_1$ states are modified by the crystal field, 
and pointed out the dominant contribution of the Van Vleck component of the susceptibility, $\chi_{VV}$. 
An explicit calculation of the susceptibility from energy levels obtained by optical spectroscopy measurements\cite{chang64, buijs87} was performed in this work. However, the resulting $\chi$  turned out  to be smaller than the experimental value, and so the remainder was 
attributed entirely to the exchange coupling among Eu$^{3+}$ ions\cite{huang69}.

This viewpoint  has been challenged on the basis of measurements of $\chi$ in  solid solutions 
of Eu$_2$O$_3$ into A$_2$O$_3$ (A=Y,Lu,Sc)\cite{antic97,concas08,grill70,kern71},
where Eu atoms are diluted. 
As the number of Eu nearest neighbors is reduced,
the total interatomic exchange interaction should 
decrease and become negligigle at small concentration of Eu, and the total susceptibility should decrease along with it. The experimental 
susceptibility (per mole of Eu),  however,
does not decrease in the full range of investigation 
(i.e. up to 10 \% of Eu$_2$O$_3$ into A$_2$O$_3$\cite{antic97,concas08,grill70,kern71}), suggesting a negligible role of the exchange interaction.
This decrease of the exchange component on dilution might be due to compensation caused by change in 
the crystal field splitting of Eu sites in A$_2$O$_3$ in comparison with Eu$_2$O$_3$; 
this explanation, however,  is not supported by optical measurements of the energy 
levels\cite{buijs87,malta95,zych02}. 
Yet another explanation has been proposed based on the distribution of Eu atoms 
in the two available sites (symmetry S$_6$ and C$_2$) of the cubic bixbyite structure 
of these oxides\cite{hanic84}. The decrease of the exchange component on dilution might be compensated by a 
preferential occupation of the S$_6$ site in the solid solution; the Van Vleck 
susceptibility of this site is larger because the  $^7$F$_1$ levels are lower in 
energy in comparison with the C$_2$ site. This preferential occupation, however, was not found in 
X-ray diffraction\cite{antic97}, and M{\"o}ssbauer spectroscopic studies\cite{concas08,concas03}. 

A further criticism to viewpoint that the exchange coupling guides the physics of suceptibility also 
comes from the calculations 
of the Van Vleck susceptibility performed by Caro et al.\cite{caro86}. These authors determine the crystalline field 
parameters starting from the experimental values of the energies of the 
excited states, and the matrix elements of the Van Vleck susceptibility were calculated 
using this potential and the atomic functions. The susceptibility calculated accordingly
agrees well with the experimental one\cite{borovik56}, validating a picture in which the exchange coupling 
should be negligible. We should mention here, however, that the optical data used by Caro et al. 
have been subsequently corrected by more recent measurements\cite{buijs87}.

Therefore, the question of whether the excess of susceptibility may be entirely attributed 
to the exchange coupling is still open. Absence of long range magnetic order
does not allow direct evaluation of the exchange coupling constants starting 
from experimental data and so a theoretical determination of these parameters by means of \emph{ab-initio}
calculations is highly desirable for understanding the behaviour of susceptibility in the material.
Remarkably, to the best of our knowledge, there is no experimental evidence in the literature of 
exchange coupling between $J = 0$ ions making the "excess" of susceptibility of 
Eu$_2$O$_3$ a novel puzzle where {\em ab-initio} calculations have the potential of
providing interesting evidence towards the solution of this puzzle.

From the theoretical point of view, rare earth compounds  represent a challenge for modern
electronic structure calculations. Their multiplet structure cannot be explained on the basis of
a single Slater determinant. The traditional density functional theory (DFT) methods with
local/semi-local approximations to the exchnage correlation functionals 
fail to describe their correlated nature and
result into a qualitatively wrong picture with  flat $f$ bands accumulated 
around the Fermi level ($E_{F}$).  To overcome these problems the LDA+$U$, and the 
self-interaction corrected local density  approximation (SIC-LDA)  methods have 
been widely used in the past. Focusing entirely Eu$^{3+}$ compounds, LDA+$U$ 
calculations were performed by Johannes and Pickett\cite{johannes} on EuN and EuP,  
and by Deniszczyk\emph{et al.}\cite{euf3} on EuF$_{3}$ and EuCo$_{2}$X$_{2}$ (X=Si,Ge). 
SIC-LDA calculations were performed on several rare earth oxides by Petit  \emph{et al.}\cite{petit,petit07}.
All these works show that the correct physics in these materials can be treated within the 
LDA+$U$ (SIC-LDA) method. 
Hence in the present work we compute the electronic and magnetic structure of Eu$_2$O$_3$ using the 
LDA+$U$ approach. Eu being one of the heavy rare earths, the use of an all-electron method is almost essential-- 
in the present work we emply full-potential linearized augmented plane wave (FLAPW) method\cite{flapw}
implemented within the Elk code\cite{elk}. 
From the corresponding results we obtain very small values for the exchange coupling  constants in cubic Eu$_2$O$_3$, 
which imply a negligible contribution of exchange to the magnetic susceptibility.
 
\section{Computational Details}
\label{meth}
For the cubic structure a {\bf k}-point mesh of $2\times2\times2$  
is used for the Brillouin zone (BZ) integration;  given the large dimension of this system
(40 atoms per unit cell) and the insulating character of the compound, 
this choice is a reasonable compromise between accuracy and computational load.
For the (smaller unit cell) hexagonal structure, a mesh of $4\times4\times4$ is used.
Spin-orbit coupling has been included in all self-consistent calculations.
We have used the fully-localised-limit of the LDA+$U$ method\cite{LDA+U}. The values of
$U$ for Eu $f$-states and $J$ are chosen to be 7 eV and 0.75 eV respectively\cite{kunes}.

\section{Crystal and Magnetic Structure}
\label{structure}
\begin{figure}
\includegraphics[scale=0.3]{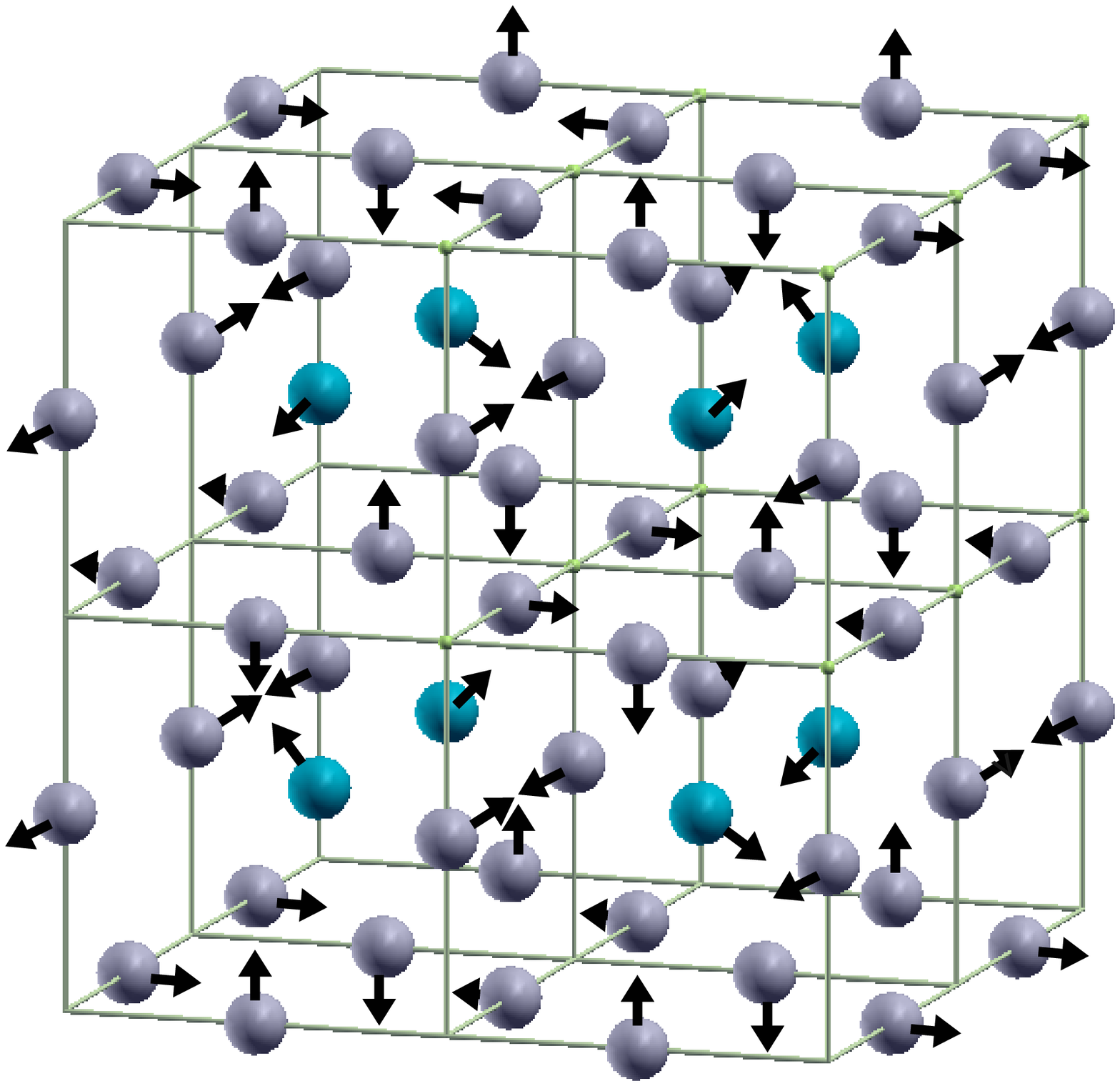}
\includegraphics[scale=0.3]{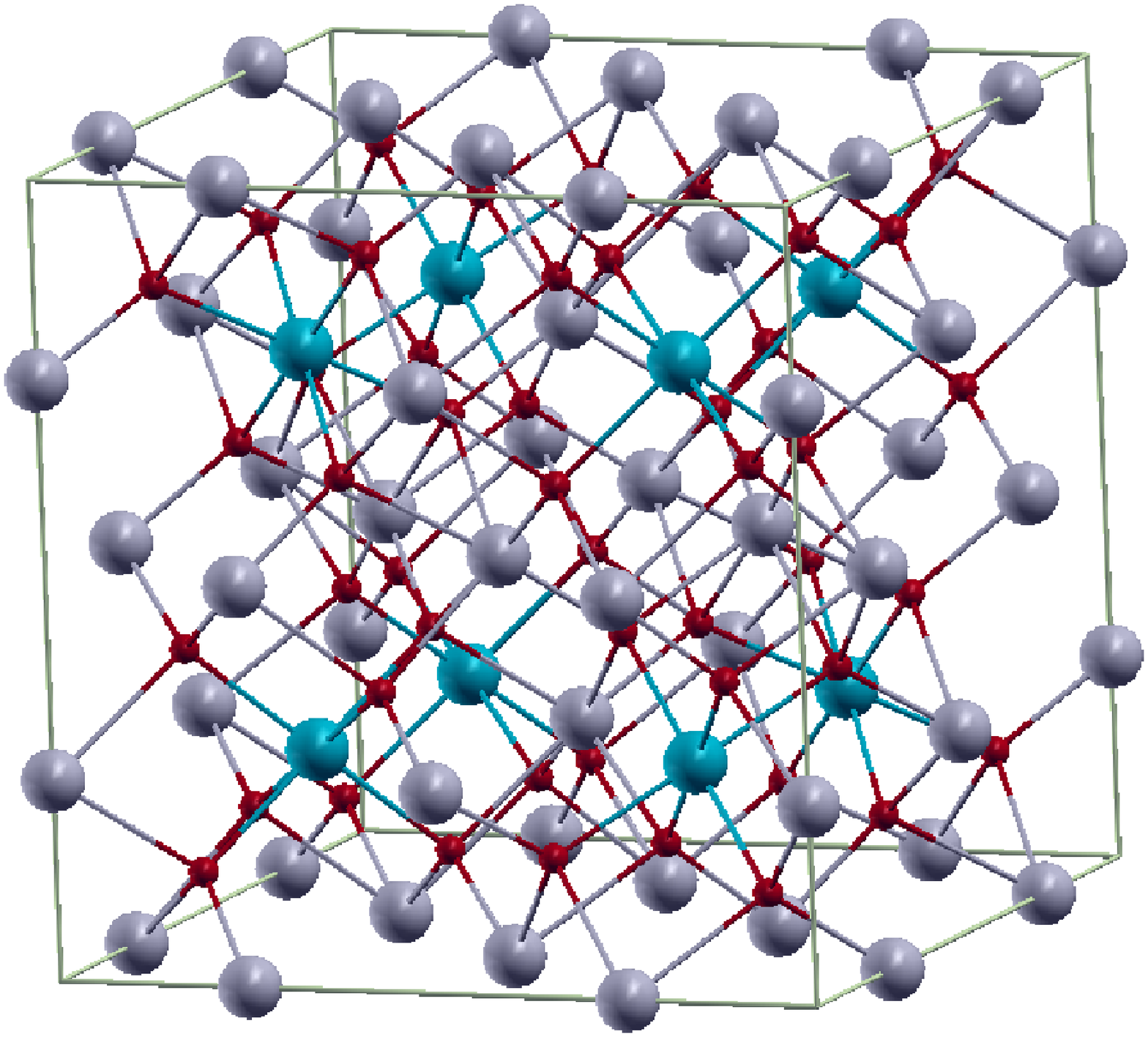}
\includegraphics[scale=0.3]{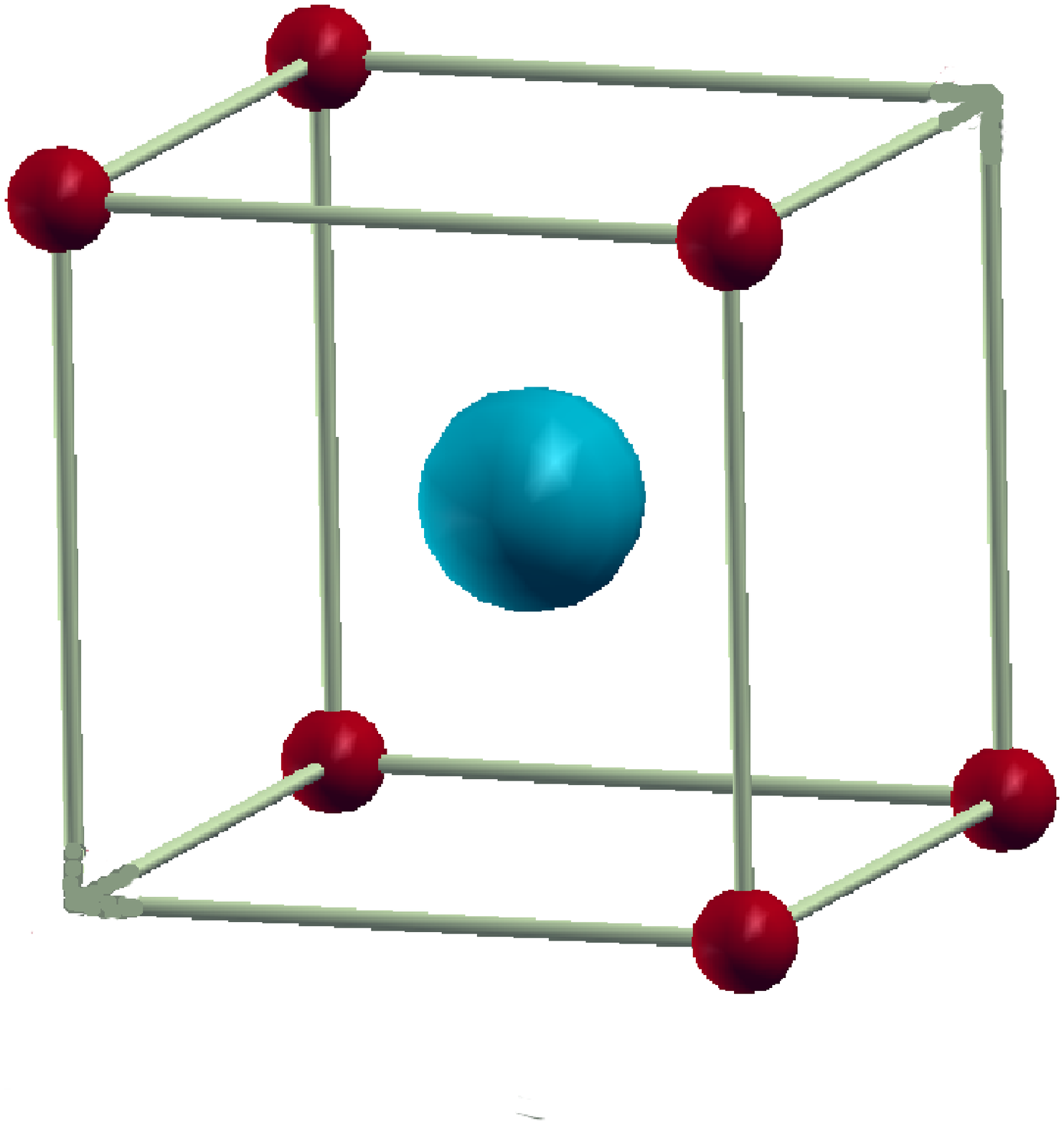}
\includegraphics[scale=0.3]{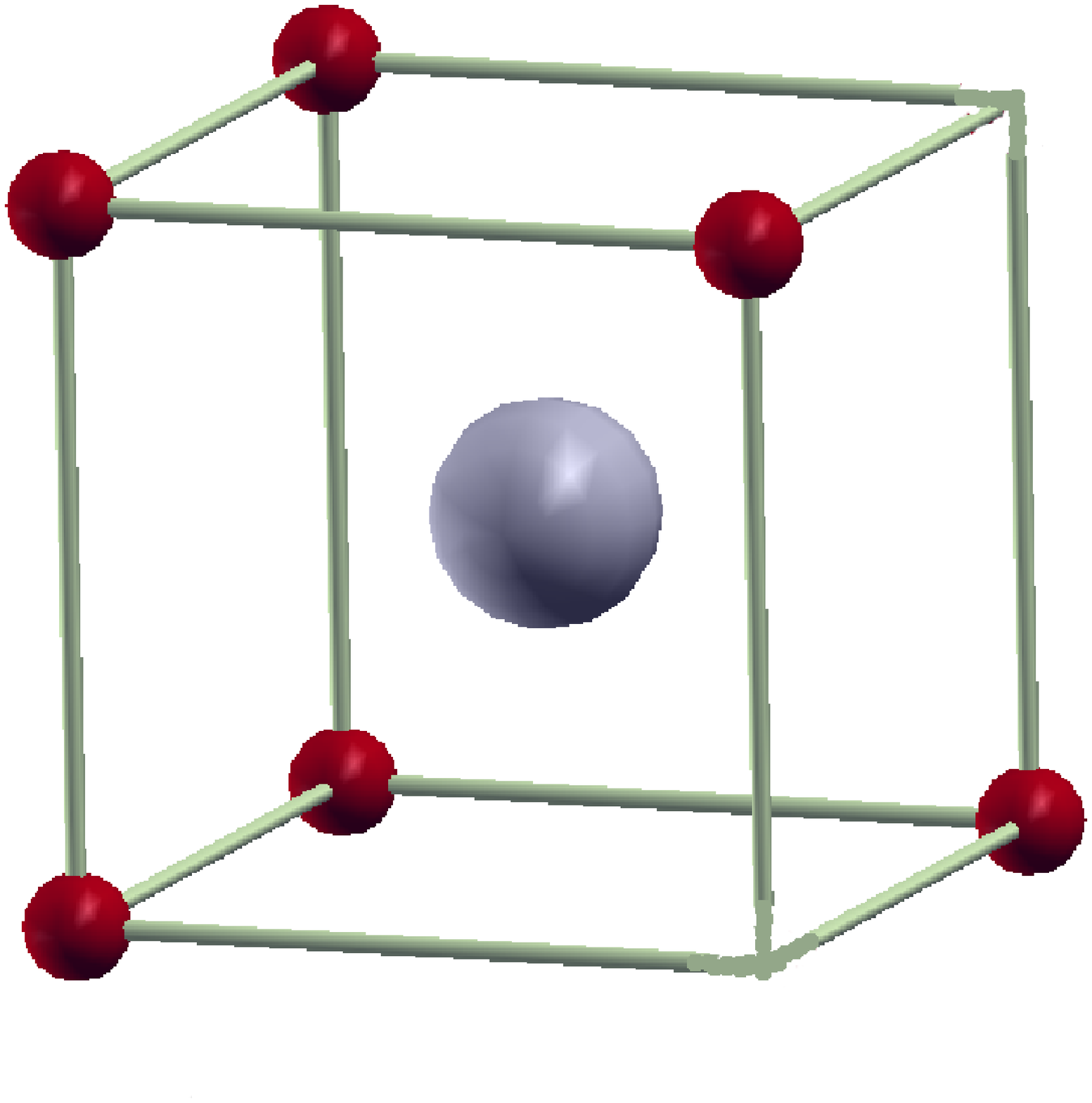}
\caption{(Color online) (a): Spin structure of the Er$_2$O$_3$ determined using neutron diffraction\cite{moon68} method. 
(b): Full structure of cubic Eu$_2$O$_3$.
(c) O coordination around Eu $S_6$ and(d) O coordination around Eu $C_2$ sites. 
}
\label{fig:structure}
\end{figure}

Eu$_2$O$_3$ crystallizes in the cubic bixbyite structure (shown in Fig.~\ref{fig:structure}) 
with space group $T^{7}_{h}$ ($Ia\bar3$); it corresponds to a $bcc$ lattice with a primitive cell 
of 40 atoms\cite{wyckoff}. Eu atoms occupy the two inequivalent Wyckoff sites 8a 
(site symmetry S$_6$) and 24d (site symmetry C$_2$), with a distorted octahedral coordination of O atoms (in the 48e sites). 
The experimental lattice constant for Eu$_2$O$_3$ is $a=10.859$~\AA \cite{hanic84}. 
The bixbyite structure is derived from  a defective cubic fluorite  structure, which corresponds 
to a simple cubic lattice with 10 atoms and half value of the lattice constant. Eu atoms occupy 
the positions of a $fcc$ lattice; the defective structure is obtained by removing two O atoms 
at 1/4 and 3/4 of the body diagonal.  
Due to the large number of atoms in the unit cell  in the bixbyite structure, Ref.~\onlinecite{petit} 
used this smaller, defective fluorite unit cell. In order to 
compare our results with this work, we have also performed calculations within the defective fluorite unit cell.

In order to evaluate the magnetic exchange coupling constant, the 
band calculations need to be performed with different 
configurations of the Eu spins. Experimentally, Eu$_2$O$_3$ does not have a long 
range magnetic order; thus as a starting point we have used
the spin configuration of isostructural Er$_2$O$_3$ compound (see Fig.~\ref{fig:structure}).
This spin configuration also turns out to be a stable for Eu$_2$O$_3$ in our calculations.
However, in this Er$_2$O$_3$-like configuration the sum of the scalar products 
of the spin of one ion with its nearest neighbors (NN) spins  is always zero, which does not allow
for a determination of the NN coupling within a Heisenberg model. In order to obtain a positive 
sum of scalar products one half of the spins of the C$_2$ sites have been reversed
relative to the Er$_2$O$_3$-like configuration such that all the spins point along the positive 
direction of the cartesian $x$, $y$ or $z$ axes. 
In order to compute the exchange couplings we have also studied other ordered phases--
to obtain a ferromagnetic (FM) like state where the direction of the S$_6$ sites is chosen to maximize 
the number of positive components, which may be written as (1,1,1), (1,1,-1), 
(-1,1,1) and (1,-1,1) in the appropriate units; we obtain in this way a positive sum of scalar products for 
a S$_6$ or C$_2$ europium with its twelve NN. In order to obtain an antiferromagnetic (AFM) like state, 
the S$_6$ spin direction is reversed; the sum of products is then negative.

\begin{figure}
\includegraphics[scale=0.3]{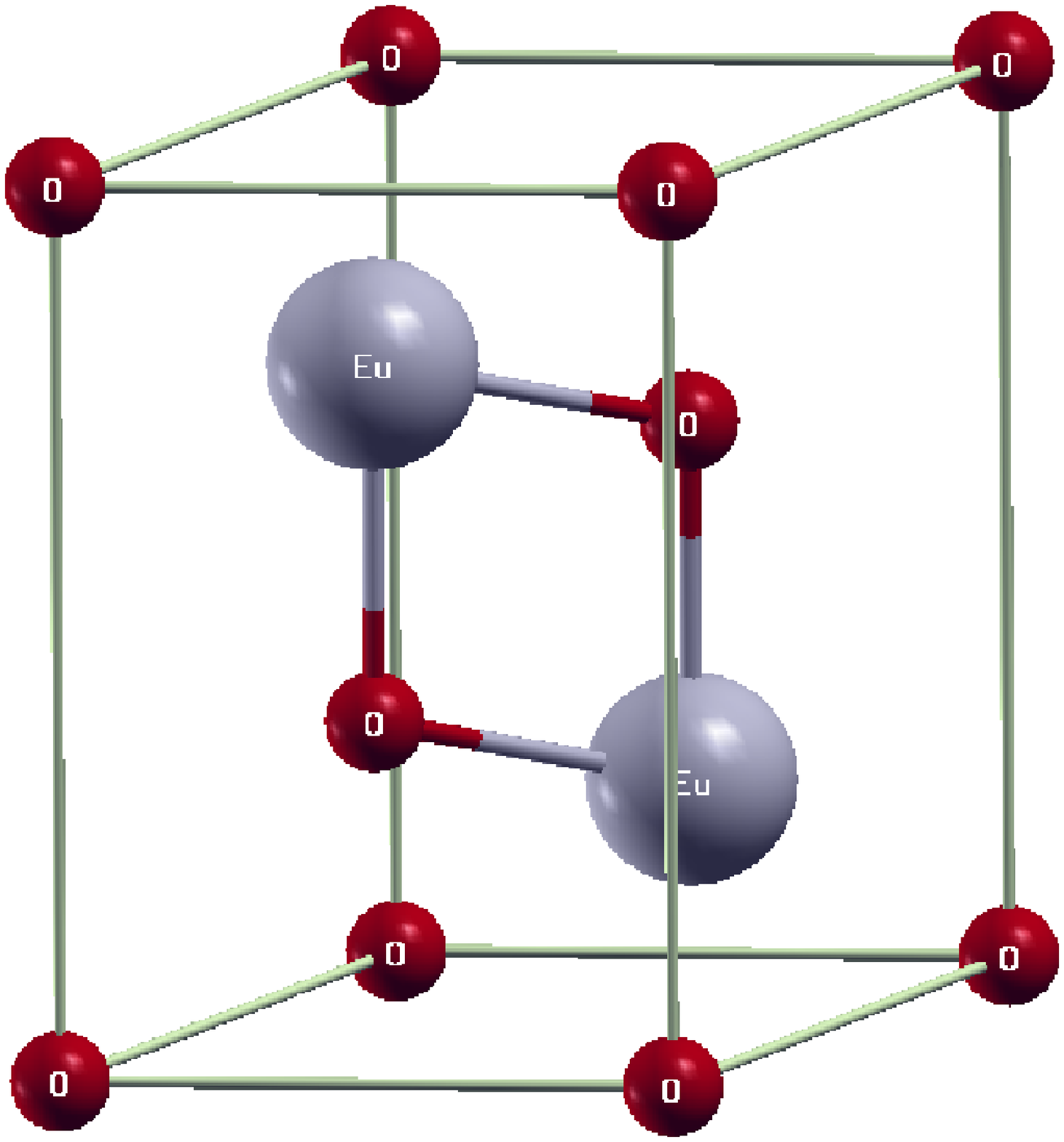}
\includegraphics[scale=0.3]{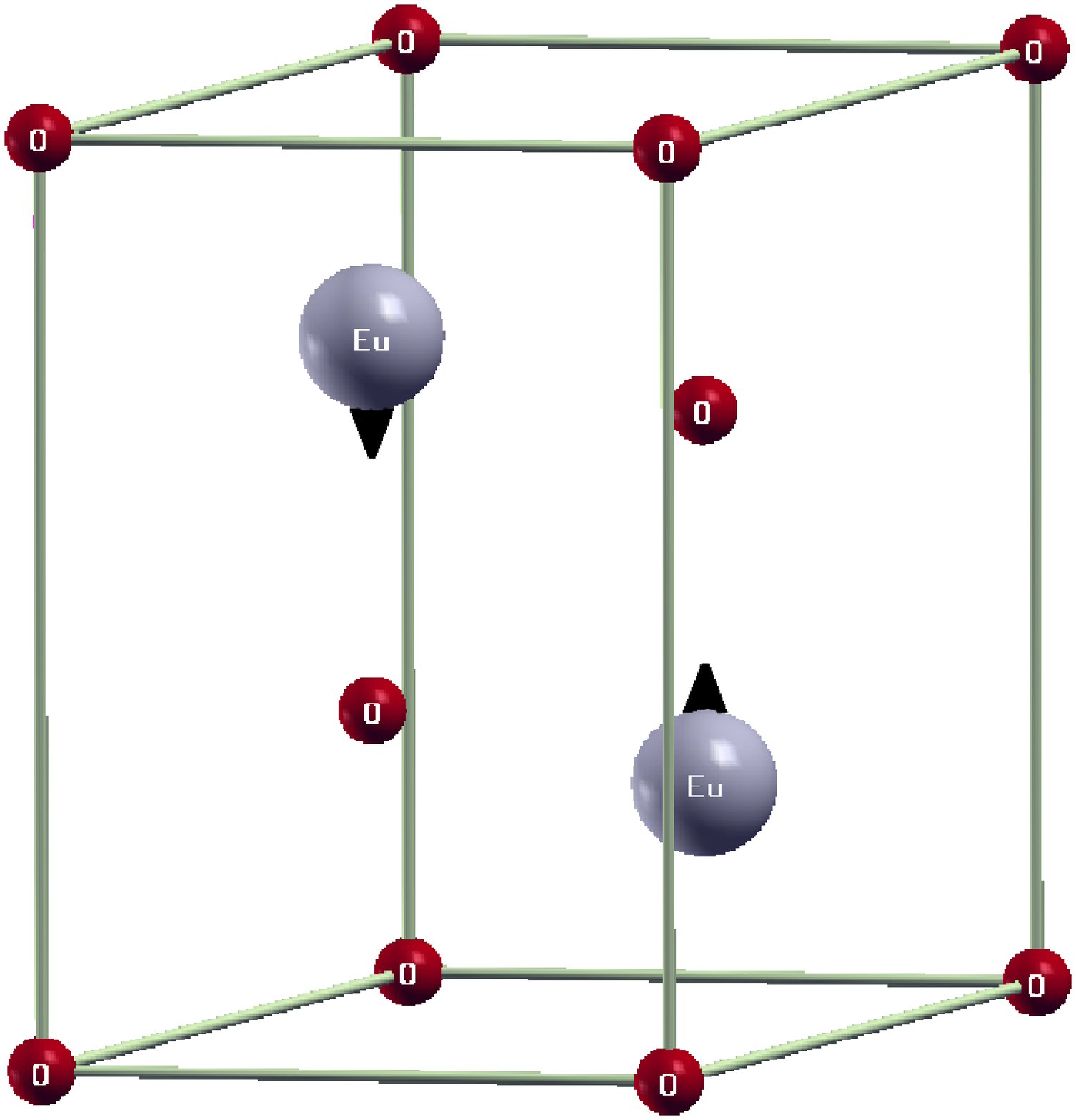}
\caption{(Color online) Left panel: high pressure hexagonal structure of Eu$_2$O$_3$.  Right panel:
antiferromagnetic configuration of the hexagonal phase.}
\label{fig:structure2}
\end{figure}

Eu$_2$O$_3$ undergoes a structural transition under pressure from the cubic to the hexagonal 
phase (space group $P\bar3m1$)\cite{jiang}  (see Fig.~\ref{fig:structure2}), which has 5 atoms 
per primitive cell (Eu atoms on the Wyckoff positions $2d$ and O atoms on the 
$2d$ and $1a$ positions)\cite{wyckoff}. The transition 
begins at 5 GPa and completes at 13 GPa. At this pressure, the lattice parameters are $a=3.738$~\AA~ and 
$c=5.632$~\AA\cite{jiang}. For this structure, the band calculations are performed with 
FM as well as AFM spin configuration; in the AFM 
configuration (shown in Fig.~\ref{fig:structure2}), the two cations of the primitive cell have antiparallel spin orientation.
 
\section{Results and discussion}
\label{resul}
 Before calculating the electronic structure a full structural relaxation is performed and the results are 
compared with the available experimental data.
Due to the large size of the bixbyte unit cell, these calculations for the cubic phase have been performed 
using the defective fluorite cell with a FM configuration. 
The volume dependence of the energy was fitted to the third order Birch-Murnaghan 
equation of state\cite{birch}. We obtain 
an equilibrium lattice constant $a_0 = 5.311$~\AA, 2 \% smaller than the experimental value
(consistent with the typical error in LDA or LDA+$U$ calculations).
The calculated bulk modulus $B_0 = 140$ GPa, is also in good agreement with the experimental 
value of $B_{0, exp} = 145\pm2$ GPa\cite{jiang} (measured in the bixbyte phase).

As mentioned earlier, the first problem in the study of this system is the choice of spin configuration, since  
 experiments do not report any magnetic order in this $J=0$ system. 
 We therefore assume as a starting (reference) configuration 
 the experimental magnetic structure of isostructural Er$_2$O$_3$. 
In this configuration, spin moments are not collinear, and the two 
crystallographic sites have totally different directions\cite{moon68}: 
while the $S_{6}$ moments are directed along the diagonal axes, $C_{2}$ sites direct their 
spin along the Cartesian axes (see Fig.~\ref{fig:structure}).

\begin{figure}
\includegraphics[scale=0.35]{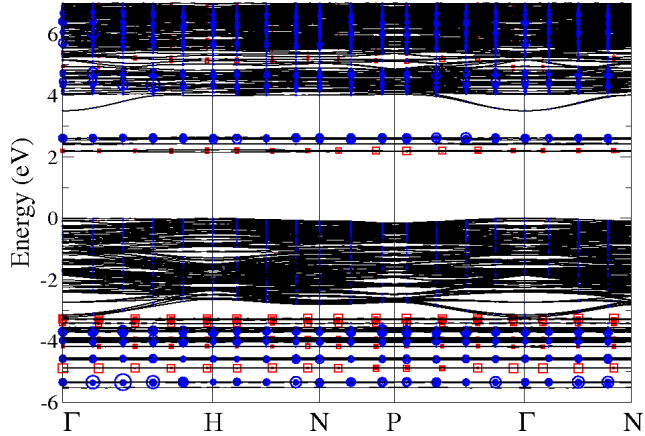}
\caption{(Color online) Band structures of cubic Eu$_2$O$_3$
along the symmetry directions of the Brillouin zone. Eu $4f$ character is denoted by red squares and blue
circles for the S$_{6}$ and C$_{2}$ sites respectively.}
\label{fig:bande}
\end{figure}

The band structure of cubic Eu$_2$O$_3$ with the Er$_2$O$_3$-like spin configuration is shown in Fig.~\ref{fig:bande}.  
The red squares and blue circles in Fig.~\ref{fig:bande} represent
 the importance of the contribution from Eu $4f$ states, for sites $S_{6}$ and $C_{2}$ respectively.
 The general features of Eu$_2$O$_3$ bands can be readily explained as follows: 
 the O $p$ states are responsible for the  bands located in the energy region between $\approx -3$ eV and the Fermi 
 level (which is the zero of energy in all our plots). The Eu$^{3+}$ ions are in a $4f^{6}$configuration, 
 which leaves an
 empty $4f$  orbital per atom in the majority spin channel. Minority $4f$ states, on the other hand,
 are completely empty and form the group of bands from +4 to +8 eV. 
 Occupied majority spin $4f$ states form the group of 24 bands sitting from -5.5 to -3 eV and 
 their splitting into four separated groups is similar to the situation in Eu pnictides\cite{johannes}.
 This splitting in the case of Eu pnictides was interpreted as an effect of the intra-atomic
 anisotropic exchange-- a test calculations based on LDA resulted into a single narrow $f$-manifold 
indicating that splitting cannot originate from crystal field effects, making the complex exchange 
effect in this open shell system the most likely explaination. 
 The large distance between (equal spin) filled  and
 empty counterparts derives from the large $U$ value for localized $4f$ states. The empty 
 majority spin states lie around 2.3 eV above  $E_{F}$. This electronic structure therefore 
 reflects the electronic configuration of the Eu$^{3+}$ ions. 
 All the $4f $ bands have a negligible dispersion ($\approx 0.2 $ eV), consistent with the
 localized nature of these orbitals.

\begin{figure}
\includegraphics[scale=0.3]{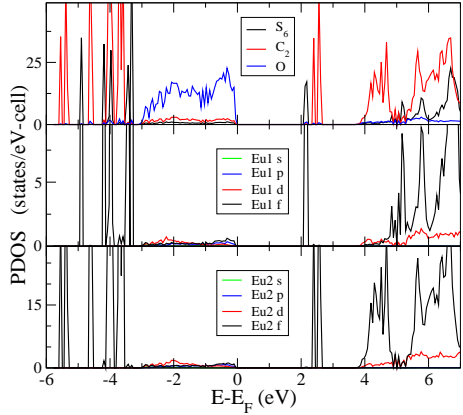}
\caption{(Color online) Orbital projected DOS of cubic Eu$_2$O$_3$ per primitive cell in the Er$_2$O$_3$-like configuration. The Eu1 and Eu2 labels indicate the S$_6$ and C$_2$ sites respectively.}
\label{fig:dos-orb}
\end{figure}

For better understanding of the electronic states presented in Figs. \ref{fig:dos-orb}
 are the  densities of states (DOS) of cubic Eu$_{2}$O$_{3}$ in the Er$_2$O$_3$-like 
 spin configuration. 
 Figs. \ref{fig:dos-orb} confirms the picture given above. In particular, the states from -3 eV to $E_{F}$ 
derive from O states, and hybridize only weakly with Eu states. 
Besides $4f$ states, Eu's most prominent contribution to the electronic structure is the $5d$ states, 
mostly located in the conduction bands region above 4 eV. 
The narrow peaks from -5.5 to -3 eV and the structures
 above 4 eV are derived from filled majority and empty minority  Eu $4f$ states respectively. 
  The peaks in the region $\approx 2-2.5$ eV above 
 $E_{F}$ are predominantly Eu states (in particular, empty majority spin $4f$ states), with a very small contribution from 
 oxygen orbitals. The marginal Eu $4f$- O $2p$ hybridization indicates towards a small value of the NN exchange coupling.
 
Importance of the crystal field (CF) effects in this compound has been a subject of discussion in past\cite{huang69,buijs87}. 
Optical measurements have been interpreted assuming that the $S_{6}$ and $C_{2}$ sites have a very different CF splitting. 
 Our results support this  interpretation (see Fig. \ref{fig:dos-orb}) and show that the $f$ states  
behave differently for the two sites-- the lower symmetry of the $C_{2}$ site results in a larger subdivision of the $4f$ 
peaks with occupied $f$ states shifted 0.5 eV lower in energy compared to the occupied $f$-states with $S_{6}$ site symmetry.
The unoccupied $f$-states show a similar behaviour with $C_{2}$ site symmetry states moved higher in energy with respect to $S_{6}$ site 
projected $f$-states. This leads to  $C_{2}$ site having a much larger energy gap between occupied
and empty majority spin states as compared to $S_{6}$ site, which in turn implies different density matrices for the two sites.
This difference can be related to the different expectation values of orbital momentum (to be discussed later). 

\begin{figure}
\includegraphics[scale=0.3]{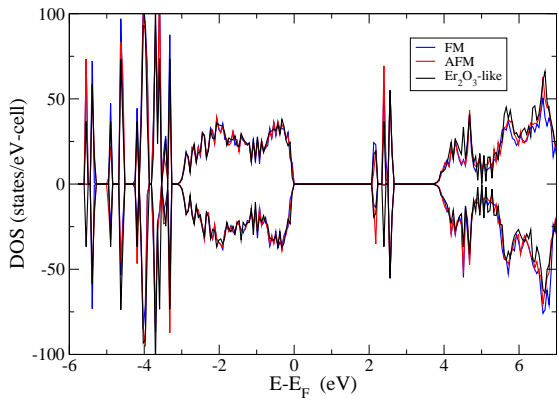}
\caption{(Color online) Total DOS of cubic Eu$_2$O$_3$  in the Er$_2$O$_3$-like configuration, and in the two (FM and AFM)
configurations described in the text.}
\label{fig:dos-spins}
\end{figure}

In order to understand the exchange interaction in Eu$_2$O$_3$  we studied two further spin 
 arrangements, refered to as FM and AFM configurations (we refer to Sect.~\ref{structure} for their description).
 In Fig.~\ref{fig:dos-spins} we compare the DOS for Eu$_2$O$_3$ in the FM and AFM configurations.
 It is clear that spin ordering affects the Eu states marginally and the corresponding DOS are 
 almost identical. These results have consequences
 on the exchange parameters as the total energy difference between these two configurations is very small. 

We have also calculated the expectation values of the quantum operators 
$\mathbf{L}$, $\mathbf{S}$ and $\mathbf{J}$ for the two different sites.  The quantization axes are different for the two sites:
in the case of $C_{2}$ it points along the Cartesian axes, and in the $S_{6}$ site along the main cube diagonal. 
In units of $\hbar$ we obtain for the $S_{6}$ site $S_{\alpha}=2.77$,
$L_{\alpha}=-1.56$ and $J_{\alpha}=1.21$. 
Our value of  $S_{\alpha}$ differs from the expected value $S_{\alpha}=3$ for the free ion probably because of the use of 
muffin-tin sphere for computing integrals, and also because of an incomplete spin polarization.
In the case of the $C_{2}$ site, on the other hand, we get very different results: 
the components along the corresponding local quantization axes $\alpha'$ are
$S_{\alpha'}=2.41$, $L_{\alpha'}=-2.34$ and $J_{\alpha'}=0.07$. 
In other words, the $J=0$ ground state of Eu seems to be reproduced to a much larger extent.  
Again (consistent with experiments) we find very different behaviors for the two sites.
We should mention that our small value of $L_{\alpha}$ for the $S_{6}$ site is similar to the
value $L_{\alpha}=-1.5$ obtained in EuN by Johannes and Pickett\cite{johannes}, who
ascribed this result to  an overquenching of angular momentum  caused by an overestimated 
crystal field effect within DFT.

At this point it is interesting to compare the electronic structures of  Eu$_2$O$_3$ and EuN\cite{johannes}  
(which crystallizes in the rocksalt  structure and correspond to an Eu$^{3+}$ configuration).
The  general features of the band structures are  quite similar, a part from obvious differences associated with 
 the different ligand (the center of gravity of O $2p$ states lies slightly deeper in energy). 
 A qualitative difference exists in the properties of Eu$_2$O$_3$ and EuN: EuN is metallic in nature and this metallic 
character arises from partially occupied dispersed  Eu $5d$  bands, crossing the empty majority $4f$ bands. 
On the other hand in case of Eu$_2$O$_3$  similar dispersed bands with Eu $5d$ contribution exist but they also have 
a relevant interstitial character and lie above the empty majority $4f$ bands; as a consequence, 
Eu$_2$O$_3$ turns out to be an insulator.
It is interesting to notice that both in our calculations for the simplified structure suggested by Petit et al. \cite{petit}
and in their SIC calculations these bands cross $E_{F}$
and in contrast with experiments gives a spurious metallic character to Eu$_2$O$_3$.  
The origin of the difference in character between the bixbyte and the simplified structures lies in the relaxation 
around the vacancies, which is not allowed by symmetry in the latter.
In fact, an inspection of the charge corresponding to these dispersed bands in the model structure shows that these 
bands derive from quantum states localized in the O vacancy sites. 

Under pressure Eu$_{2}$O$_{3}$  undergoes a structural phase transition to a hexagonal phase  
(see Fig.~ \ref{fig:structure2}).  Due to the very large computational load involved, 
the calculation of the transition pressure is out of the scope of the present investigation.
However, we minimized the total energy of Eu$_{2}$O$_{3}$  in the 
hexagonal phase.
The fitted energy minimum is at $V_0 = 67.62$~\AA$^3$, and  $B_0 = 180$ GPa.  
The calculated bulk modulus compares well with the experimental value 
$B_{0, exp} = 151\pm6$ GPa\cite{jiang}. 
As for the equilibrium volume, the experimental hexagonal volume is $V_{exp} 
= 68.15$~\AA$^3$ when the transition is completed (at 13.1 GPa)\cite{jiang}. 
Using the theoretical bulk modulus, we  arrive at an equilibrium 
volume at 13.1 GPa equal to 62.7 \AA$^3$, which underestimates the experimental lattice constant 
by about 2.8~\% (which is typical of LDA based functionals). 

\begin{figure}
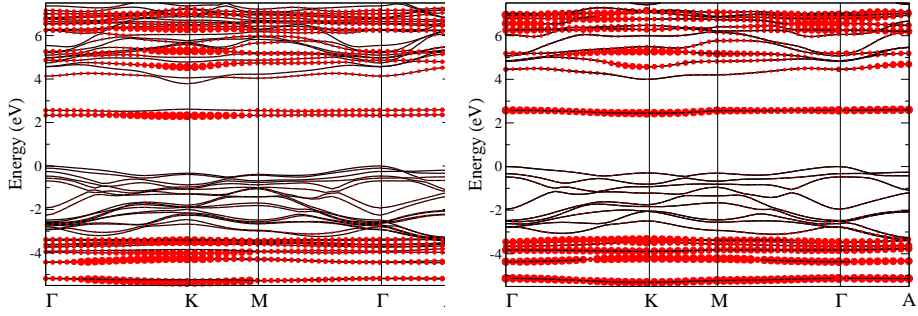

\includegraphics[scale=0.25]{figure6a.eps}
\includegraphics[scale=0.25]{figure6b.eps}
\caption{(Color online) Band structures of hexagonal Eu$_2$O$_3$
in the primitive cell. Left panel: ferromagnetic (FM) state. Right
panel: antiferromagnetic (AFM) state. Eu $4f$ character is denoted by red
circles. 
}
\label{fig:bande-hex}
\end{figure}

In order to understand how pressure modifies the electronic structure of this compound,
we have studied its properties in the hexagonal phase, at the lattice parameters corresponding to $P=13.1$ GPa. 
The band structure of the hexagonal Eu$_2$O$_3$ is shown in Fig.~\ref{fig:bande-hex} 
for both FM and AFM configurations. The general features of the bands are similar to the 
cubic phase;  hexagonal Eu$_2$O$_3$ is a semiconductor, 
with a gap of 2.3 eV between valence and unoccupied $4f$ band, and a gap of 4.1 eV between valence 
and conduction band. 
This semiconducting behaviour is in agreement with the experimental optical and transport properties of 
Eu$_2$O$_3$. 
In contrast to the cubic phase, in the hexagonal phase the dispersed band with the minimum 
around 3.5 eV at the $\Gamma$ point are not present.
This difference may be related to the fact that the hexagonal structure has 
no vacancies in the atomic O positions ruling out the possibility 
of having bands with a large interstitial character.

\begin{figure}
\includegraphics[scale=0.3]{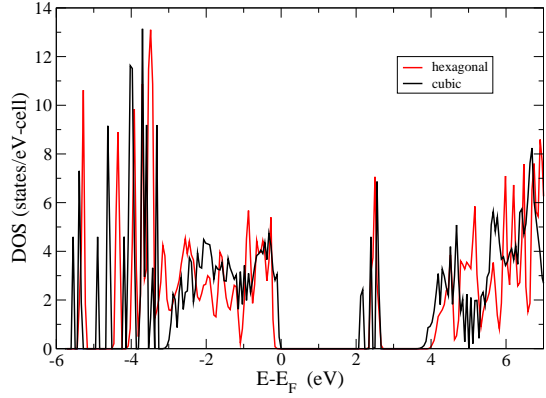}
\caption{(Color online) Total DOS  of cubic and hexagonal Eu$_2$O$_3$ in the FM state.  \\ }
\label{fig:dos-eu-cub-hex}
\end{figure}

The total DOS  of cubic and hexagonal Eu$_2$O$_3$ are presented in 
Fig.~\ref{fig:dos-eu-cub-hex}. While the structure of levels is similar, the
O $2p$ states have larger bandwdith and overlap with the occupied
Eu $4f $ manifold. This increase of the O $2p$ bandwitdh is clearly a consequence of the compression
O atoms experience under pressure.
Also, the exchange-related splitting of the $4f$ levels is different in the cubic and hexagonal phases.

Some results of our calculations on cubic Eu$_2$O$_3$ may be compared with experimental values. The gap between the valence and the conduction band ($\Delta E_{v->c}$) has been obtained by optical measurements; the calculated value $\Delta E_{v->c}=3.50$ eV is close to the experimental one $\Delta E_{opt}=4.3\pm0.3$ eV\cite{prokofiev}. The gap between the valence and the empty majority spin Eu $f$-band has been  determined by the experimental curve of conductivity vs. temperature; the calculated value $\Delta E_{v->f}=2.18$ eV is in reasonable agreement with the experimental value of $\Delta E_{cond}=1.84$ eV\cite{lal}. 

Our calculations allow us to calculate the electron density on the nucleus, which is relevant in the context of the M{\"o}ssbauer spectroscopy of $^{151}$Eu in cubic Eu$_2$O$_3$.  
The electron density on the nucleus may be evaluated in terms of the difference relative to tetragonal EuF$_3$, which is the reference compound. Using the experimental isomer shift $\delta=1.03\pm0.01$ mm/s\cite{concas03} and the differential nuclear radius of Ref.~\onlinecite{shenoy}, we find an experimental density difference to be $\Delta \rho(0)=3.18\pm0.03$ a$_0^{-3}$; the calculated  value 
for  the dominating  C$_2$ site is $\Delta \rho(0)=3.17$ a$_0^{-3}$ which is in very good agreement
with the experimental value.

\section{Exchange interactions}
\label{exchange}

The exchange interaction in ions with $L \neq 0 $ is characterized by the dependence of the exchange integral 
on orbital orientation. According to Van Vleck and  Huang \cite{vanvleck69} this effect 
gives rise to an "anisotropic exchange", resulting from the dependence of orbital charge density on the direction. 
The coupling between the ions $i$ and $j$ may be described by the 
exchange potential\cite{vanvleck69}
\begin{equation} V_{ex} = -\sum_{i=1}^{n_i}\sum_{j=1}^{n_j}\sum_{\mu,\mu'}\sum_{\nu=-\mu}^{\mu}
\sum_{\nu'=-\mu'}^{\mu'} a_{\mu\nu,\mu'\nu'}Y_{\mu\nu}^*(\mathbf{l}_i)Y_{\mu'\nu'}(\mathbf{l}_j)
(\frac{1}{2}+2\mathbf{s}_i\cdot\mathbf{s}_j) \label{eq:1}\end{equation}
where $Y_{\mu\nu}$ are the tesseral harmonic operators equivalent to Ref. \onlinecite{vanvleck69}; $n_i$ is 
the number of electrons in incomplete shell of the ion $i$, $\mathbf{s}_i$ is the electron spin and 
$\mu = \mu' = 6$ for Eu$^{3+}$.
As demonstrated in Ref. \onlinecite{vanvleck69}, the exchange coupling for Eu$^{3+}$ ions in their 
ground state and in cubic compounds, may be written in the standard form
\begin{equation} V_{ex} = -2a_{eff}^{(ij)}\mathbf{S}_i\cdot\mathbf{S}_j \label{eq:2}\end{equation}
where $\mathbf{S}_i$ is the spin of the ion $i$ and $a_{eff}^{(ij)}$ are the effective exchange 
constants. In other words, we can use the standard form of the isotropic coupling, 
to deal with an anisotropic exchange. 
Therefore, it is possible to determine the $a_{eff}^{(ij)}$ constants by calculating the 
energies of different spin configurations. These constants allow to determine the effect of the exchange 
coupling on the magnetic susceptibility.

Our results show that there is no significant total energy difference between 
the FM-like and the AFM-like phases (for their definition see section \ref{structure}), 
with an energy convergence parameter 
$\epsilon \sim 3 \times 10^{-4}$ eV and a total energy $E_t \sim 5 \times 10^{6}$ eV. The 
maximum possible strenght of the exchange coupling consistent with these results corresponds 
to an energy difference per primitive cell $\Delta E_t \sim 2\epsilon$ which leads to a 
difference per Eu atom of $\Delta E \sim 0.04$ meV. In this work we consider only the exchange 
coupling of the Eu$^{3+}$ ion with its twelve NN, since it is expected to be the largest.
In fact, in the bixbyite structure, NN exchange is mediated by the oxygen atom, which is 
not the case for  next NN Eu atoms. 
With this assumption, the  effective exchange constants refer only to the coupling of NN Eu ions.
We also assume that all the NN pairs have the same $a_{eff}^{(ij)} = a_{eff}$. 
Starting from the energy 
difference $\Delta E \sim 0.04$ meV and the spin values given by our calculations, we obtain
that the upper bound of the effective exchange constant to be $a_{eff} \sim 0.002$ meV.

Due to lack of long range magnetic order\cite{morrish}, the value of $a_{eff}$ cannot be verified 
by direct comparison with experimental magnetic data.
However, this interaction may give a contribution to the magnetic susceptibility.
According to Ref.~\onlinecite{huang69},  the magnetic susceptibility $\chi$ may be written as
\begin{equation} \chi = \chi_{dia} + \chi_{p} = \chi_{dia} + \chi_{VV} + \chi_{ex} \label{eq:5}\end{equation} 
where the paramagnetic susceptibility $\chi_p$, obtained by subtraction of the diamagnetic core
component $\chi_{dia}$, is the sum of the Van Vleck contribution $\chi_{VV}$ and of the exchange contribution
$\chi_{ex}$ \cite{huang69}.
At $T$ = 0 K,
\begin{equation} \chi_{VV} = \frac{8N\mu_B^2}{3K_{B}} \sum_{k=1}^3 \frac{1}{E_{1k}} \label{eq:6}\end{equation}
\begin{equation} \chi_{ex} = \frac{128N\mu_B^2A_{eff}}{E_1(E_1-16A_{eff})} \label{eq:7}\end{equation}
where N is the number of atoms, $\mu_B$ is the Bohr's magneton, $K_{B}$ is the Boltzann's 
constant, $E_{1k}$ are 
the energies of the  triplet  state $^7$F$_1$ centered around the energy $E_1$, and  
\begin{equation} A_{eff} = \sum_{j=1}^{12} a_{eff}^{(ij)} = 12a_{eff} \label{eq:8}\end{equation}
However, the contribution of the S$_6$ and C$_2$ sites must be evaluated separately, because
the energies of the $^7$F$_1$ state are different. Since the multiplet energies
are not accessible from our one-particle calculations, we calculate the Van Vleck
contribution by using the experimental optical energies for the two sites given by
Ref. \onlinecite{buijs87};  we get $\chi_{VV}^{(exp)} = 7.72 \times 10^{-3}$ cm$^3$/mol 
of Eu (in CGS units). If we subtract this contribution from the experimental
value of the paramagnetic susceptibility $\chi_{p}^{(exp)} = 9.1 \times 10^{-3}$ cm$^3$/mol of Eu 
(in CGS units) of Ref.~\onlinecite{concas08}, we obtain the value of the excess susceptibility
$\Delta\chi = \chi_p - \chi_{VV} = 1.38 \times 10^{-3}$ cm$^3$/mol of Eu. 

In accordance with Van Vleck and Huang\cite{huang69,vanvleck69} if we assume that  
this excess susceptibility is due to the exchange contribution only ($\Delta\chi = \chi_{ex}$)
and that $A_{eff}$ is equal for both sites, we can estimate the experimental value of 
$A_{eff} = 0.41$ meV.
By using the maximum value of $a_{eff}$ obtained by our band calculations, we obtain 
a theoretical upper bound for the constant to be $A_{eff} \sim 0.02$ meV. 
Our results, therefore,  
 lead to the conclusion that in cubic Eu$_2$O$_3$ the difference between
the experimental paramagnetic  susceptibility and the Van Vleck contribution cannot be due to
the contribution of the exchange interaction,
in contrast with conclusions of Van Vleck and Huang\cite{huang69,vanvleck69}.
It is consistent, on the other hand, with the fact that the experimental susceptibility 
(per mole of Eu) does not decrease in the solid solutions of Eu$_{2}$O$_{3}$ into 
A$_{2}$O$_{3}$.  
The point of view of Van Vleck and Huang has also been criticized before by Caro \emph{et al.}\cite{caro86},
who performed a calculation of the Van Vleck contribution $\chi_{VV}$ including the matrix elements
among atomic wavefunctions of the Eu $4f$ states.  They obtained a resulting $\chi_{p}$ in good agreement
with experiment, without invoking any contribution from the exchange interaction. 
Our calculations are also consistent with these results.

The investigation of the exchange interaction has also been performed for the hexagonal phase 
of Eu$_2$O$_3$ under pressure. In this case, Eq.~\ref{eq:2} is not applicable, because the structure 
of the compound is not cubic; therefore, it is not possible to determine the exchange constants by 
total energy differences. In contrast to the cubic phase in the case of the hexagonal structure the 
application of a weak magnetic field along the $c$ axis gives collinear spins.  In the AFM configuration  
(described in Section \ref{structure}) the cation spins have parallel orientation in planes perpendicular 
to the magnetic field direction ($z$ axis), with antiparallel orientation between planes. 
The Eu atom has three neighbors at 3.50 \AA~ in the lower plane, three at 3.59 \AA~ in the upper plane, 
and four neighbors at a distance of 3.74 \AA\  in the same plane\cite{wyckoff, jiang}. The energy 
difference per Eu atom is $E_{AFM} - E_{FM} = -3.63$ meV. Therefore in the hexagonal 
structure we obtain a small but sizable AFM exchange interaction between the Eu ion and its six NN. 
Interestingly the energy difference is of the same order of magnitude as that found by Johannes 
and Pickett for EuN and EuP\cite{johannes}. 

The exchange interaction is about 100 times stronger in the hexagonal phase than in the cubic one. 
As a first guess, one might attribute this difference to 
the shorter Eu-Eu distance in the hexagonal phase 
($d_{Eu-Eu}$ is 3.50-3.74~\AA\ in the hexagonal phase and 3.84~\AA\ in the cubic one).
To verify this possibility, we performed test calculations 
in the hexagonal structure, changing the lattice constant so as to match the  interatomic Eu-Eu distance of 
the cubic phase. The corresponding exchange coupling did not decrease sufficiently to support this view.
The rational for the different couplings may probably be found  in the different local O coordination around Eu ions, as 
the oxygen orbitals mediate the Eu-Eu interaction: 
in the hexagonal structure there are two O atoms binding two NN Eu ions, while there is only one intermediate 
O atom in the cubic one. An analysis of the partial densities of states in the two structures supports this view, 
with a quite large hybridization of Eu $4f$ and O orbitals in the hexagonal phase.
Interestingly we also observe that the phase with stronger exchange coupling has a collinear spin structure, while the structure
with weaker interaction is characterized by non-collinear spins.

\section{Conclusion}
\label{concl}
In conclusion, we studied the electronic structure of cubic bixbyte Eu$_{2}$O$_{3}$ and its high pressure hexagonal 
phase by mean of full potential linearized augmented plane wave calculations, within the LDA+$U$ method.
In both phases the filled O $2p$ and empty majority spin Eu $4f$ states are separated by a $\approx 2.1$ eV gap, 
while minority Eu $4f$ states start around 4 eV
above $E_{F}$, in agreement with the optical and transport measurements\cite{prokofiev,lal}.

From the comparison of FM and AFM calculations we infer that the
interatomic exchange interaction is negligible in cubic Eu$_2$O$_3$, and so is its effect on the susceptibility. 
This is consistent with the experimental observation of a constant susceptibility 
(per mole of Eu) of solid solutions of Eu$_{2}$O$_{3}$ 
into A$_{2}$O$_{3}$ (A=Y, Lu, Sc). Our calculations are in contrast with the point of view of 
Van Vleck and Huang\cite{huang69,vanvleck69}, according to which exchange is needed to explain the behavior 
of the magnetic susceptibility. Our results are consistent with later calculations\cite{caro86} which could explain 
the experimental susceptibility by including matrix element in the Van Vleck contribution.
In the hexagonal phase we observe the presence of a small but sizable AFM exchange interaction 
between the Eu ion and its six nearest neighbors. 
Therefore, in this phase we could have a non negligible contribution of exchange to the magnetic susceptibility, 
but for now no experimental measurements are available for comparison.

\begin{acknowledgments}
SM thanks M.D. Johannes for useful discussions. Work partially supported by a computing grant at CINECA (Italy). 
\end{acknowledgments}

\end{document}